\documentclass[sigconf]{acmart}

\makeatletter
\def\@ACM@checkaffil{
    \if@ACM@instpresent\else
    \ClassWarningNoLine{\@classname}{No institution present for an affiliation}%
    \fi
    \if@ACM@citypresent\else
    \ClassWarningNoLine{\@classname}{No city present for an affiliation}%
    \fi
    \if@ACM@countrypresent\else
        \ClassWarningNoLine{\@classname}{No country present for an affiliation}%
    \fi
}
\makeatother

\renewcommand\footnotetextcopyrightpermission[1]{} 
\usepackage{fancyhdr}
\AtBeginDocument{%
  }

\usepackage{multirow}
\usepackage{pifont}
\usepackage{eucal}
\usepackage{amsfonts}
\usepackage{amsthm}
\usepackage{amsmath}
\newtheorem{definition}{Definition}

\usepackage[ruled,linesnumbered]{algorithm2e}
\usepackage{url}
\usepackage{graphicx}
\usepackage{subfig}
\usepackage{makecell}
\usepackage{appendix}
\usepackage{booktabs}

\setcopyright{acmcopyright}
\copyrightyear{2018}
\acmYear{2018}
\acmDOI{XXXXXXX.XXXXXXX}

\acmPrice{15.00}
\acmISBN{978-1-4503-XXXX-X/18/06}

\begin{document}

\title{Extracting Cloud-based Model with Prior Knowledge}



\author{Shiqian Zhao}
\affiliation{%
  \institution{Nanyang Technological University}
}
\email{shiqian.zhao@ntu.edu.sg}

\author{Kangjie Chen}
\affiliation{%
  \institution{Nanyang Technological University}
}
\email{kangjie001@ntu.edu.sg}

\author{Meng Hao}
\affiliation{%
  \institution{University of Electronic Science and Technology of China}
}
\email{menghao@std.uestc.edu.cn}

\author{Jian Zhang}
\affiliation{%
  \institution{Nanyang Technological University}
}
\email{jian_zhang@ntu.edu.sg}

\author{Guowen Xu}
\affiliation{%
  \institution{City University of Hong Kong}
}
\email{guowen.xu@foxmail.com}

\author{Hongwei Li}
\affiliation{%
  \institution{University of Electronic Science and Technology of China}
}
\email{hongweili@uestc.edu.cn}

\author{Tianwei Zhang}
\affiliation{%
  \institution{Nanyang Technological University}
}
\email{tianwei.zhang@ntu.edu.sg}

\renewcommand{\shortauthors}{Zhao et al.}

\begin{abstract}
Machine Learning-as-a-Service, a pay-as-you-go business pattern, is widely accepted by third-party users and developers. However, the open inference APIs may be utilized by malicious customers to conduct model extraction attacks, i.e., attackers can replicate a cloud-based black-box model merely via querying exquisitely sampled or crafted examples. 
Existing model extraction attacks mainly depend on the posterior knowledge (i.e., predictions of query samples) from Oracle. 
Thus, they either require high query overhead to simulate the decision boundary, or suffer from generalization errors and overfitting problems due to query budget limitations. 

To mitigate it, this work proposes an efficient model extraction attack based on prior knowledge for the \textit{first} time. 
The insight is that prior knowledge of unlabeled proxy datasets such as intrinsic property and natural relationships is conducive to the search for the decision boundary (e.g., informative samples). 
Specifically, we leverage self-supervised learning including autoencoder and contrastive learning to pre-compile the prior knowledge of the proxy dataset into the feature extractor of the substitute model. 
Then we adopt entropy to measure and sample the most informative examples to query the target model. 
Our design leverages both prior and posterior knowledge to extract the model and thus eliminates generalizability errors and overfitting problems. 
We conduct extensive experiments on open APIs like Traffic Recognition, Flower Recognition, Moderation Recognition, and NSFW Recognition from real-world platforms, Azure and Clarifai. 
The experimental results demonstrate the effectiveness and efficiency of our attack. For example, our attack achieves 95.1\% fidelity with merely 1.8K queries (cost 2.16\$) on the NSFW Recognition API from the Clarifai platform. 
Also, the adversarial examples generated with our substitute model have better transferability than others, which reveals that our scheme is more conducive to downstream attacks.

\end{abstract}

\maketitle

\section{Introduction}
\label{section: introduction}
Machine Learning as a Service (MLaaS) is an emerging business paradigm that enables individual and cell corporations to enjoy high-performance models. 
In pay-as-you-go MLaaS solutions, companies or platforms train large-scale models and provide their Application Programming Interfaces (APIs) to the public, and customers obtain the prediction results from APIs at a reasonable cost. For instance, ChatGPT-4 charges 0.09\$ on average for every 1k tokens for the 32K context model~\cite{OpenAIPricing} (see Table \ref{tab: prediction API platforms and pricing} for more platforms). 
Despite its popularity, this pattern reveals such a risk: due to the wide availability of model interfaces, a malicious user could conduct attacks against cloud-based models via APIs, such as membership inference~\cite{shokri2017membership, long2018understanding, salem2018ml}, backdoor~\cite{liu2017trojaning, yao2019latent} and model extraction~\cite{tramer2016stealing, jagielski2020high, yu2020cloudleak, wang2018stealing, chandrasekaran2020exploring}. 

\begin{figure}[t]
\centering
\setlength{\abovecaptionskip}{0.cm}
\includegraphics[width=0.8\linewidth,scale=1.0]{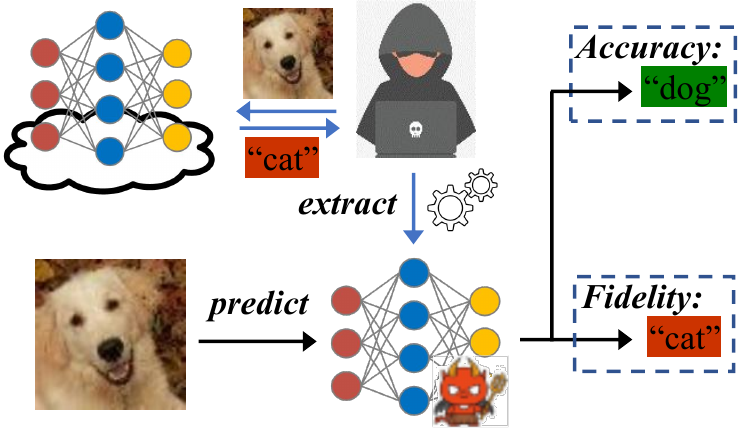}
\caption{
The workflow of model extraction attacks to steal accuracy and fidelity. The purpose of accuracy stealing is to improve the accuracy of the substitute model, while fidelity stealing targets to be consistent with the original model on both correct and wrong decisions.
}
\label{fig: The process of ME}
\vspace{-12pt}
\end{figure}

Among all the attacks, model extraction attack (MEA)~\cite{tramer2016stealing, jagielski2020high, yu2020cloudleak, wang2018stealing, chandrasekaran2020exploring} is one of the most influential ones. As shown in Figure \ref{fig: The process of ME}, the attacker can extract a cloud-based model via interactions with APIs merely: (i) the attacker queries the victim API with carefully crafted or sampled samples from the proxy dataset, and obtains the predictions from the victim model; (ii) the attacker trains a substitute model based on the obtained knowledge for high accuracy and/or fidelity. 
Apart from the harm caused to intellectual property (IP), MEAs also facilitate downstream attacks, such as adversarial attacks~\cite{papernot2017practical, juuti2019prada, oh2019towards, wang2021delving} and membership inference attacks~\cite{liu2022membership}, which capitalize on the similarity and transferability between the target and substitute models. Therefore, it is crucial to study the potential risks resulting from MEAs, so as to protect the IP of the cloud-based services~\cite{sha2022can, liu2022stolenencoder, yu2020cloudleak, papernot2017practical}.

\begin{table*}[t]
\centering
\setlength{\belowcaptionskip}{5pt}
\setlength{\abovecaptionskip}{0.cm}
\begin{tabular}{ccccccc}
\toprule[1pt]
\\[-9pt]
                  Platform&Product&Customization&Label&Posteriors&Monetize&Pricing Per K Queries \\[2pt] \toprule[1pt]
\\[-11pt]
\multirow{3}{*}{Microsoft\cite{MicrosoftAPI}} &Entity Recognition&\ding{52}&\ding{52}&Top-k&\ding{52}&\$0.3 \\
                  &Anomaly Detector&\ding{52}&\ding{52}&Top-1&\ding{52}&\$0.314 \\
                  &Celebrity Recognition&\ding{52}&\ding{52}&Top-k&\ding{52}&\$1 \\[5pt]
\multirow{2}{*}{Google\cite{GoogleAPI}} &Object Detection&\ding{56}&\ding{52}&Top-k  &\ding{52}&\$1.25 \\
                  &Logo Detection&\ding{56}&\ding{52}&Top-k&\ding{52}&\$1.2 \\[5pt]
 Face$^{++}$\cite{FaceAPI}  &Human Body Recognition&\ding{56}&\ding{52}&Top-k&\ding{52}&\$1.5 \\ 
 Clarifai\cite{ClarifaiAPI}&Image Classification&\ding{56}&\ding{52}&Top-k&\ding{52}&\$1.2 \\[1pt] \bottomrule[1pt]
\end{tabular}
\caption{The details of MLaaS products provided by various platforms. Almost all existing products are monetized and some of them support customization.}
\label{tab: prediction API platforms and pricing}
\end{table*}

Existing MEAs mainly adopt a sample-query-train paradigm to extract the information of the victim model's decision boundary, and strike a trade-off between high-performance and low-cost requirements. 
These works can be divided into two types: generation-based and search-based strategies. 
The generation-based MEAs like Data Free~\cite{truong2021data, zhou2020dast, kariyappa2021maze} utilize Generative Adversary Networks to generate query examples from noises. The limitation of this method is that it requires millions of expensive queries. Recent efforts~\cite{miura2021megex, sun2022exploring} introduce proxy datasets to address this inefficiency. 
The search-based MEAs are also based on proxy datasets. Different from the generation-based MEAs, they utilize a searching strategy to search for the most informative examples~\cite{yu2020cloudleak, papernot2017practical, juuti2019prada}. 
For example, Yu et al.~\cite{yu2020cloudleak} use adversarial active learning to iteratively craft adversary samples and query victims with them, thereby reducing the number of queries. 

Despite the impressive progress, however, there are three major drawbacks as shown in Table~\ref{overview of existing methods}. 
First, most of them are based on an ideal assumption, that is, there is a gradually converging consistency with regard to the decision boundary between the substitute model and the victim model. 
Unfortunately, under the requirement of low query overhead, it is an intractable problem due to the eternal existence of generalization error and over-fitting for few-shot training~\cite{wang2020generalizing, raghunathan2019adversarial, advani2020high}. 
Second, they all concentrate on enhancing the posterior information of queried data from victims and tend to ignore the prior knowledge from abundant unlabelled data of proxy datasets. The lack of such knowledge makes it difficult to search for the decision boundary and thus limits the performance. 
Third, existing fidelity-oriented MEAs are usually evaluated locally under some hypothesis. For example, Jagielski et al.~\cite{jagielski2020high} achieves high-fidelity model extraction when the substitute model is initialized with the same weights as the victim. 
It is still unclear how effective these attacks are when applied in real-world scenarios.

To fill these gaps, in this paper, we propose a novel model extraction attack to take the first step to steal a cloud-based model with high fidelity. The insight behind our scheme is that the posterior prediction (e.g., label and score) from Oracle is far from adequate, and the prior knowledge of unlabelled proxy datasets such as intrinsic property and natural relationships is conducive to MEAs~\cite{he2020momentum, grill2020bootstrap, chen2020simple}. Motivated by this, we leverage the prior knowledge of unlabelled proxy data and posterior knowledge of labeled query data together to alleviate the generalization error and over-fitting problem. 
However, it is non-trivial to incorporate the prior knowledge into MEAs. 
On one hand, the impact of different prior knowledge on the attack performance remains unexplored, and it is vital to select a feasible solution from various approaches. 
In this paper, we consider both generative methods and contrastive methods to comprehensively study the effect of prior knowledge on MEAs. 
On the other hand, how coordinating the prior knowledge and posterior knowledge is also challenging since they may have an inclusion or opposition relationship. 
Therefore, we propose to first utilize self-supervised learning to leach the prior knowledge from the full unlabeled proxy datasets, and then we design an active stealing framework based on entropy for sampling, which combines the label information obtained from Oracle and the knowledge learned in the previous phase. 
At the beginning of the extraction loop, the query examples are determined by the pre-compiled prior knowledge. 
This enables us to make good use of both the prior knowledge in the unlabeled proxy dataset.
As the queried data increases, the substitute model learns more posterior knowledge from the victim. Consequently, the newly sampled data increasingly relies on the posterior derived from the Oracle. 
In this way, the prior knowledge and posterior knowledge contribute together to the generalizability of query examples and the overall extraction performance.

We conduct extensive experiments and compare our scheme with state-of-the-art model extraction attacks for both Independent Identically Distributed (IID) and Out of Distribution (OOD) cases. We first evaluate the fidelity of the substitute model obtained by our method and other MEAs~\cite{correia2018copycat, pal2020activethief, orekondy2019knockoff, papernot2017practical, yu2020cloudleak}. Compared with these methods, our attack can achieve up to 35\% fidelity improvement, as well as a significant reduction of the query budget. Then we investigate how the stolen model can be utilized to implement downstream attacks such as adversarial example attacks. Compared with other methods~\cite{correia2018copycat, pal2020activethief, orekondy2019knockoff, papernot2017practical, yu2020cloudleak}, the transferability of adversarial examples generated by our substitute model surpasses the existing methods with a margin of 26\%. Also, for the same transferability, our adversarial examples require a much smaller perturbation so that are less likely to be detected. 

We also evaluate the performance of our attack in real-world MLaaS products by conducting four case studies, i.e., Traffic Recognition API, Flower Recognition API, Moderation Recognition API, and NSFW Recognition API. Our attack achieves high fidelity for stealing the NSFW Recognition model (95.1\% fidelity with 2.16\$ budget). Notably, our attack is much more effective than existing methods, which demonstrates the practicality in the real world. 

In summary, we mainly make the following contributions:
\begin{itemize}
\item Our work for the first time incorporates prior knowledge to model extraction attacks, which can better utilize proxy datasets as a starter for searching decision boundaries. 
\item We design a novel high-fidelity model extraction framework with self-supervising learning (auto-encoder and contrastive learning). Our framework can well combine the prior knowledge of the proxy dataset and the posterior knowledge from the Oracle to achieve a significant improvement in extraction fidelity. 
\item We conduct extensive experiments on popular commercial platforms and APIs such as Traffic Recognition, Flower Recognition, Moderation Recognition, and NSFW Recognition API from Microsoft and Clarifai. All the experimental results demonstrate the effectiveness of our approach.
\end{itemize}

\section{Background}
\label{sec:background}
Revolving around the data labeling, we introduce the background of this work by answering the following four questions: 1). why do we need labeled data in $\S$~\ref{supervised learning}; 2). how to solve the expensive labeling problem in $\S$~\ref{mlaas}; 3). is the current labeling paradigm secure in $\S$~\ref{subsection: model stealing}; 4). how to improve the labeling efficiency for malicious use in $\S$~\ref{subsection: representing learning}. 

\subsection{DNN Training}
\label{supervised learning}
We first introduce the general paradigm of deep neural network training. 
Let $\mathcal{D}$ = $\mathcal{X}\times\mathcal{Y}$ denotes training dataset, where $\mathcal{X} \subseteq \mathbb{R} ^{M} $ is the sample space and $\mathcal{Y} \subseteq \mathbb{R} ^{N} $ is the label space. 
The goal of deep learning is to learn a function map (also called a classifier) $f_{\theta}( \cdot)$  from the sample space $\mathcal{X}$ to the label space $\mathcal{Y}$, where $\theta$ is the learned model parameter. 
To ensure a better generalization, the classifier $f_{\theta}( \cdot)$ usually has a multi-layer structure $f_{\theta}( \cdot) = f_{K}\circ f_{K-1}\circ\cdot\cdot\cdot\circ f_{2}\circ f_{1}( \cdot)$.

In the training phase, the stochastic gradient descent (SGD) algorithm is used to iteratively optimize a loss function, e.g. the cross entropy (CE). Specifically, the training process consists of two steps: 1) computing the CE loss:
\begin{equation}
\label{cross entropy}
\mathcal{L}_{\mathsf{CE}}((\mathbf{x}, \mathbf{y});\theta)=-{\frac{1}{N}}{\sum_{i=1}^{N} \mathbf{y}_i \log \ \frac{\mathsf{exp}(f_{\theta}(\mathbf{x})_i)}{\sum_{j=1}^{N} \mathsf{exp}(f_{\theta}(\mathbf{x})_j)}},
\end{equation}
\noindent and 2) updating the model parameter with SGD:
\begin{equation}
\label{sgd}
\theta_{t+1} = \theta_{t}-{\frac{\eta}{B}}\sum _{i=1}^{B}\nabla _{\theta} \mathcal{L}_{\mathsf{CE}}(f_{\theta}(\mathbf{x}_i), \mathbf{y}_i);\theta),
\end{equation}
\noindent where $B$ is the batch size and $\eta$ is the learning rate. 

\subsection{MLaaS}
\label{mlaas}
As learned from $\S$~\ref{supervised learning}, training a well-performed model requires a mass of labeled data. 
However, the exorbitant price for labeling impedes individuals and cell corporations from obtaining their personalized model. 
Under this circumstance, Machine learning as a Service (MLaaS) gradually becomes a feasible and popular choice for everyman to enjoy high-performance large-scale deep learning models. 
Generally speaking, this pay-as-you-go pattern works in such a way: users submit query samples to platforms, which process data and return prediction results. 
To date, there are many platforms and companies offering MLaaS products as shown in Table~\ref{tab: prediction API platforms and pricing}. 

According to the ownership of the cloud-based model, MLaaS can be divided into two categories: 1). \textbf{Platforms train and own.} A mainstream MLaaS mode is that platforms train models with exclusive massive data, and open their APIs to the public for profit. Nonetheless, these APIs are often limited to common tasks such as Face Detection\footnote{https://www.faceplusplus.com/face-detection/} and General Object Recognition\footnote{https://clarifai.com/clarifai/main/models/general-image-recognition-vit}. 
2). \textbf{Platforms train but the third party owns.} For some customized requirements, a third party who is short of computing power, could train and deploy its own private model with the assistance of platforms. 
Then the third party can monetize its models and open its APIs to downstream customers. 
To sum up, although the training data is provided by a third party in the second circumstance, the training processes for both scenarios are not transparent for either the third party or users. 
Therefore, all models trained by platforms, as well as the hyper-parameters, are completely black boxes to others.

\subsection{Model Extraction Attacks}
\label{subsection: model stealing}
Tramer et al. \cite{tramer2016stealing} proposes the first concept of model extraction attack (MEA) and implements this attack on decision trees, logistic regression, and neural networks~\cite{tramer2016stealing}. 
The objective of MEA is to extract a functionally equivalent or similar substitute for the cloud-based victim model. 
In practice, model extraction attacks should meet three conditions: 1) \textit{Weak attacker}. The adversary only has black-box access to the cloud-based model and has zero knowledge about the structure and hyperparameters of this model. In other words, attackers could only obtain information about the cloud-based model via APIs. 
2) \textit{Low query overhead}. The adversary can query the target model within a small budget since the number of queries represents the economic cost for the adversary and also the risk of being detected. 
3) \textit{High functional equivalence requirement}. The substitute model stolen from the platform requires achieving a high similarity with the victim model.

Without losing generality, the mainstream MEA strategies adopt an active framework to iteratively mimic the oracle~\cite{chandrasekaran2020exploring, pal2020activethief, yu2020cloudleak}. 
More specifically, the attacker first \textit{randomly} initializes a query dataset, and then he circularly 1) queries the victim and obtains prediction results from the victim, 2) uses the labeled data to train a substitute model, and 3) determines the query data with the locally trained model for the next round. 
This circulation could repeat several times until the query budget is exhausted or termination conditions are triggered such as the functional equivalence requirement. 
According to the method of determining data for the next query, MEAs can be further classified into \textit{Generation-based} MEAs and \textit{Search-based} MEAs. 

\renewcommand{\arraystretch}{1.2}
\begin{table}[]
\setlength{\belowcaptionskip}{5pt}
\setlength{\abovecaptionskip}{0.cm}
\scalebox{0.94}{
\begin{tabular}{c|ccc}
\hline
MEAs & Black-box & Query Efficiency & Proxy Utilization \\ \hline
Tramer~\cite{tramer2016stealing}&\ding{56}&\ding{52}&\ding{56}   \\
Truong~\cite{truong2021data}&\ding{52}&\ding{56}&\ding{56}   \\
Zhou~\cite{zhou2020dast}&\ding{52}&\ding{56}&\ding{56}   \\
Correia~\cite{correia2018copycat}&\ding{52}&\ding{56}&\ding{56}   \\
Juuti~\cite{juuti2019prada}&\ding{52}&\ding{52}&\ding{56}     \\
Papernot~\cite{papernot2017practical} &\ding{52}&\ding{52}&\ding{56}      \\
Pal~\cite{pal2020activethief} & \ding{52} & \ding{52} & \ding{56}        \\
Yu~\cite{yu2020cloudleak}     & \ding{52}  &   \ding{52}  & \ding{56}  \\
Ours        &\ding{52}&\ding{52}&\ding{52}   \\ \hline
\end{tabular}
}
\caption{Overview of existing model extraction attacks.}
\label{overview of existing methods}
\vspace{-12pt}
\end{table}
\renewcommand{\arraystretch}{1}

\noindent\textbf{Generation-based MEAs.} 
The generation-based strategy assumes that adversaries do not possess any query dataset, which is also called data-free MEAs. 
In this setting, several works~\cite{zhou2020dast, kariyappa2021maze, truong2021data, yuan2022attack,zhang2022towards} utilize Generative Adversarial Networks (GAN) to craft adversarial examples located near the boundary of the target model. 
Despite this data-free advantage, these works cause a costly query overhead~\cite{miura2021megex}. 
For example, Truong et al.~\cite{truong2021data} utilize 20M queries to improve the accuracy of the substitute model from 76.8\% to 88.1\% for the CIFAR10 task. 
Therefore, this setting violates the requirement of \textit{low query overhead} in model extraction attacks. 
With consideration of the key issue of generation-based MEAs, works~\cite{miura2021megex, sun2022exploring, barbalau2020black} are proposed to address the problem of high query consumption. 
More specifically, Sun et al.~\cite{sun2022exploring} utilizes a proxy dataset to guide the generation of the synthesized dataset, and Barbalau et al.~\cite{barbalau2020black} directly adopts a generative model pre-trained with semantically unrelated proxy dataset in the general data-free framework.

\noindent\textbf{Search-based MEAs.}
On the other hand, the search-based strategy considers a more practical scenario where a positive adversary gathers a surrogate dataset and then samples the most informative examples to conduct model extraction. 
Several recent works follow this strategy and propose query-efficient attacks based on a variety of search strategies, such as Active Learning~\cite{jagielski2020high, chandrasekaran2020exploring, yu2020cloudleak, pal2020activethief, pal2019framework, psathas2022decreasing}, Reinforcement Learning~\cite{orekondy2019knockoff, chen2021stealing}, Jacobian Augmentation~\cite{papernot2017practical, juuti2019prada}. 
Despite the intuitive feasibility of this method, the crucial overfitting issue for few-shot active learning is remaining to be addressed under a constraint of \textit{low query overhead}. 
This problem is inextricable for the current mainstream framework of MEAs while considering reducing the query budget. 
Unlike previous works, this paper proposes to combine the prior knowledge of data intrinsic property and posterior information obtained from oracle to effectively alleviate this symptom. 

\subsection{Self-supervised Learning}
\label{subsection: representing learning}
As the data-hungry property of supervised learning, the model may suffer from generalization error and over-fitting problems while the training data is deficient~\cite{raghunathan2019adversarial,advani2020high}. 
In reality, however, most of the collected data is unlabelled while high-quality human annotation is particularly expensive. For example, data labeling company Scale\footnote{https://scale.com/pricing} charges 6 dollars for every image annotation. 
Under this circumstance, self-supervised Learning (SSL) is proposed to assist the model in getting rid of heavy Oracle labeling~\cite{misra2020self, hendrycks2019using, noroozi2018boosting, tung2017self}. 
The insight behind SSL is that the prior knowledge of data such as intrinsic property and natural relationships are independent of but beneficial to acquired semantic learning. 
Generally, SSL can be summarised into generative methods and contrastive methods. 
Generative methods such as Masked Autoencoder~\cite{he2022masked} mainly adopt dimensionality reduction and reconstruction to compile the most important feature of data into an encoder, as well as filter the noise contained in data~\cite{devlin2018bert, sun2019ernie}. 
Different from the reconstruction principle, contrastive methods learn the distribution of datasets by increasing the distance between similar examples and estranging distinct examples~\cite{grill2020bootstrap, chen2020simple, he2020momentum}. 

\section{Problem Formulation}
\label{sec: problem formulation}
In this section, we provide the threat model for model extraction attacks and the formal definition of fidelity-oriented MEA.

\subsection{Threat Model}
\label{subsection: threat model}

\noindent\textbf{Attack's goal.}
We consider a malicious user who tries to steal the real-world cloud-based model $\mathcal{O}$ merely via the query API. Following previous work~\cite{jagielski2020high, rakin2022deepsteal, chabanne2021side}, we target our model extraction attack to achieve high fidelity between the substitute model $\hat{\mathcal{O}}$ and the victim model $\mathcal{O}$. Different from the accuracy-oriented MEAs~\cite{tramer2016stealing, yu2020cloudleak, chandrasekaran2020exploring}, the fidelity goals ensure a consistent boundary with the victim so that it benefits the downstream attacks. 

\noindent\textbf{Attacker's capability.}
We consider a weak attacker who has black-box access to the cloud-based model in real-world scenarios. 
The attacker could only query carefully sampled examples to the victim model for inference and analyze the predictions. As we investigate the commercialized APIs, we find that most of the MLaaS products provide a detailed functionality of the model. Given these facts, in this paper, we study the vulnerability of cloud-based models to MEAs under different real-world scenarios. We characterize the attacker's ability along these two dimensions: 
\begin{itemize}
\item \textbf{Model Structure and Hyper-parameter.} As $\mathcal{O}$ is a black box for the adversary, the attacker knows nothing about the model structures of the victim. Although some APIs open their architecture choice\footnote{https://clarifai.com/clarifai/main/models/moderation-recognition}, we restrict the attacker's knowledge to make sure the attack has a fabulous generalization ability. In this paper, we adopt ResNet50 as the architecture of the substitute model. As for hyper-parameters, we use the regular setting for DNN training. For example, we use SGD as the optimization algorithm. 

\item \textbf{Surrogate Dataset}. Depending on the accessibility to the training data of the victim model, the proxy dataset $\mathcal{D}_{proxy}$ of the attacker could be categorized as two folds. a). \textbf{Independent Identically Distributed (IID)}. It is possible that $\mathcal{O}$ is trained with some datasets that are so unique that have to be used or are uploaded from users. Model owners may disclose the information of datasets out of censorship from the government or consumers. In this circumstance, the attacker could have full access to the training dataset~\cite{jagielski2020high, yu2020cloudleak}. b). \textbf{Out of Distribution (OOD)}. More commonly, the MLaaS providers keep the training datasets secret. However, to facilitate the transaction with customers, platforms usually offer a clear description of the model's functionality, i.e., the concepts of categories and the scope of service. Therefore, in practice, a positive adversary could collect task-related OOD data from the public. 
\end{itemize}

\subsection{Fidelity Model Extraction}
\label{subsection: threat model}
While most of the existing works~\cite{tramer2016stealing, yu2020cloudleak, chandrasekaran2020exploring} are keen on improving the extracted model's accuracy, Jagielski et al.~\cite{jagielski2020high} and the following works~\cite{rakin2022deepsteal, chabanne2021side} explore fidelity-orientated MEAs and illustrate the critical advantages, especially when using model extraction to launch downstream attacks such as black-box adversarial example~\cite{papernot2017practical, juuti2019prada}, membership inference~\cite{liu2022membership}, and attribute inference attacks~\cite{zhao2021feasibility}. In this paper, we also concentrate on the fidelity goal, which can be formally defined as:

\begin{equation}
\label{fidelity metric}
\mathcal{F} ={\frac{1}{|\mathcal{X}|}}\sum_{\mathbf{x} \in \mathcal{X}} \operatorname{Argmax}(\mathcal{O}(\mathbf{x}))=\operatorname{Argmax}(\hat{\mathcal{O}}(\mathbf{x})),
\end{equation}

where $\mathcal{O}(\mathbf{x})$ and $\hat{\mathcal{O}}(\mathbf{x})$ denote the posterior probabilities of the victim and substitute models, respectively. Given the adversary’s goal and fidelity metric, we formally define our model extraction attack as follows:
\begin{definition}[$(t, q)$-Model Extraction Attack (MEA)]
\label{definition:mea}
Given a black-box access to a cloud-based target model $\mathcal{O}$, a query budget of $q$, a fidelity goal of $t$, and a validate dataset $\mathcal{D}_{val}$, $(t, q)$-MEA aims to obtain a substitute model $\hat{\mathcal{O}}$ within $q$ queries to $\mathcal{O}$ such that 
\begin{equation}
    \mathcal{F}(\hat{\mathcal{O}}(\mathcal{D}_{val}),\mathcal{O}(\mathcal{D}_{val})) \geq t. 
\end{equation}
where $\mathcal{F}$ is the fidelity metric as above. 
\end{definition}

\section{Our Attack}
\label{subsection: our attack}

\begin{figure*}[t]
\centering
\includegraphics[width=0.9\textwidth,scale=1.00]{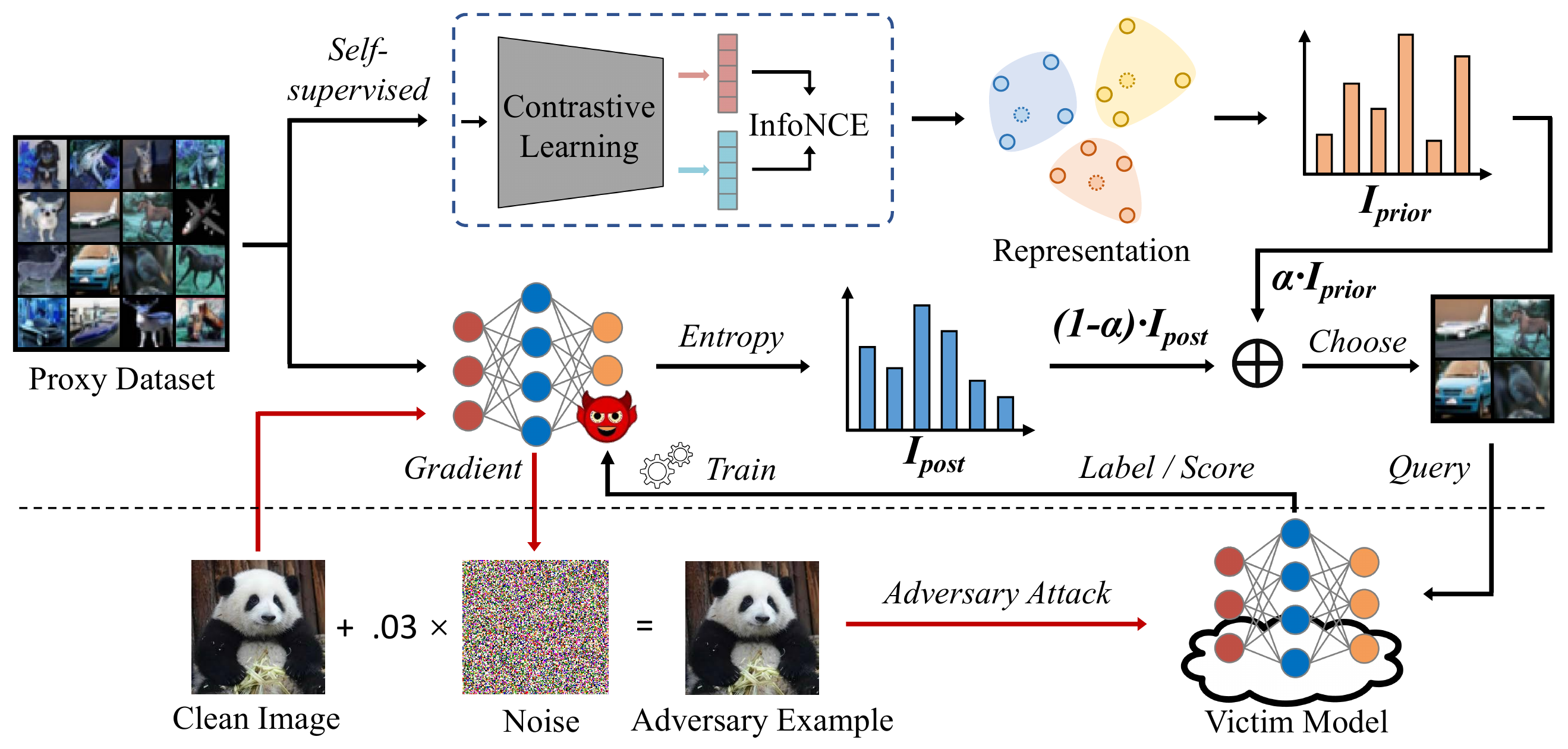}
\caption{The framework of our attack. It mainly includes two parts: 1). embedding the prior knowledge into the encoder and clustering the samples according to their representations; 2). imitating the victim model with its output and using a combination of prior and posterior knowledge to determine the most informative samples. After obtaining a well-preformed substitute model, we could utilize it to conduct downstream attacks.}
\label{framework}
\end{figure*}

\subsection{Problem Analysis}
\label{subsection: problem analysis}
\noindent\textbf{Active Learning for MEA.} In active learning-based MEA, given an Oracle (well-performed cloud-based model), active learning helps alleviate the heavy human annotation burden by sampling the most informative samples $\mathcal{D}_{query}$ from unlabelled datasets $\mathcal{D}_{proxy}$ to tag. Note that the most informative examples are those with the highest uncertainty for \textit{locally trained substitute model $\hat{\mathcal{O}}$}. Therefore, when the posterior (annotated data) is little, unfortunately, the decision boundary of the substitute model could have a colossal variance from that of the victim model $\mathcal{O}$. The current solutions aim to increase the information contained within query samples while ignoring the abundant prior knowledge included in the large-scale unlabelled proxy dataset. 
To counter such symptoms, we leverage the new emerging technique and propose a novel prior-guided model extraction scheme with entropy information. 

\noindent\textbf{Prior-guided MEA.} The prior knowledge in context is the information existing within the surrogate dataset of the attacker. 
Intuitively, the samples with semblable semantic annotation should have similar activation in the representation space, and vice versa. 
Such prior knowledge inspires the development of Self-supervision Learning, which advocates focusing on the data itself. 
Motivated by the great achievement made by SSL, we introduce the prior knowledge learning to the MEA task and propose a novel entropy-based MEA scheme. 
The key idea of our method is that in the initial phase of a MEA attack, the informative samples can be determined by the prior knowledge of the proxy dataset; and then as the query budget increases, the decision-making power gradually shifts to the posterior. 
Through the corporation and competition of prior and posterior knowledge, the sampled points could have the best transferability to the victim model.

\subsection{Prior Knowledge Learning}
\label{subsection: prior knowledge learning}
To leverage a learning-based method to mount our model extraction attack, we consider compiling the prior knowledge into the feature extractor of the substitute model. 
More specifically, we utilize diverse self-supervised learning algorithms to explore the inherent property and natural relationship within the proxy dataset. 
Theoretically, the paradigm of SSL can learn a good representation space that is conducive to downstream supervised learning tasks. 
In this paper, we consider five prior knowledge including four SSL methods:

\noindent\textbf{Random Sampling (RS):} We first consider a simple solution, random sampling, where an attacker stochastically selects examples from his proxy dataset without replacement. Then he queries the APIs with sampled points all at once and constructs a training dataset using the query results for the substitute model training. The advantages of RS are two-fold: on one hand, it guarantees an identical distribution from the query dataset to the surrogate dataset, and on the other hand, its sampling decision does not require interaction with Oracle and thus is hard to detect. 

\noindent\textbf{Basic Autoencoder (BAE):} For high-dimensional data, it is common sense that there are too many trashy or even negative features existing in raw data or embeddings. To mitigate their influence, Ballard et al.~\cite{ballard1987modular} propose to utilize an autoencoder to pre-train the artificial neural networks. After decades of development, a typical autoencoder usually includes an encoder and a decoder in series. As a typical paradigm, autoencoders attempt to filter out unimportant features by dimension reduction and reconstructing primitive images as formulation: 
\begin{equation}
\begin{aligned}
\label{ae}
minim&ize \ d(x,f_{dec}(f_{enc}(x))),\\
\mathrm{s.t.}& \ |f_{enc}(x)| \ll |x|
\end{aligned}
\end{equation}
where $f_{enc}, f_{dec}$ are the encoder and decoder of AE, respectively, and $d$ denotes the similarity measurement. To control the degree of filtering quantificationally, we can change the output dimensionality of $f_{enc}(x)$. The prior knowledge behind BAE is that the raw data has inessential features which are deleterious for model training. 

\begin{algorithm}[t]
\caption{MEA Based on Prior Knowledge.}
\label{alg:PriorMS}
\KwIn{Proxy Dataset: $\mathcal{D}_{proxy}$; Query Budget: $B$; Victim: $\mathcal{O}$; Interaction Times: \textit{Itera};}
\KwOut{Substitute Model: $\hat{\mathcal{O}}$;}
\BlankLine
$f_{enc} = SSL(\mathcal{D}_{proxy})$

$\hat{\mathcal{O}} \leftarrow f_{enc} + fc$

$\mathcal{D}^{ite}_{query}\leftarrow$ Select $B/Itera$ samples with the highest $Entropy(\hat{\mathcal{O}}(\mathcal{D}_{proxy}))$, $ite=0$

\While{$ite \leq Itera$}{
$\mathcal{D}^{ite}_{queried}=CallAPI(\mathcal{D}^{ite}_{query})$

$\mathcal{D}^{ite}_{queried}=\mathcal{D}^{ite-1}_{queried}\cup \mathcal{D}^{ite}_{queried}$

$\hat{\mathcal{O}} \leftarrow f_{enc} + fc$

Train $\hat{\mathcal{O}}$ with $\mathcal{D}^{ite}_{queried}$

$\mathcal{D}^{ite+1}_{query}\leftarrow$ Select $B/Itera$ samples with the highest $Entropy(\hat{\mathcal{O}}(\mathcal{D}_{proxy}\setminus \mathcal{D}^{ite}_{query}))$

$ite = ite+1$

}

\textbf{return} $\hat{\mathcal{O}}$

\end{algorithm}

\noindent\textbf{Denoising Autoencoder (DAE):} DAE is a popular and widely studied generative method in self-supervised learning. With the intuition that the representation of data should be robust to noise, Devlin et al.~\cite{devlin2018bert} propose to randomly mask some tokens from the input and attempt to reconstruct the previous data based on the context information inside for the NLP task. Then He generalizes this method to pre-train an encoder for an image task~\cite{he2022masked}. Formally, the optimization objective of DAE can be written as: 

\begin{equation}
\begin{aligned}
\label{dae}
minimi&ze \ d(x,f_{dec}(f_{enc}(x+\delta))),\\
\mathrm{s.t.}& \ |f_{enc}(x+\delta)| \ll |x|
\end{aligned}
\end{equation}
where $d$ denotes the $L_{2}$ distance between the recovered data and raw data. 
In the general paradise of DAE, noise $\delta$ in Function~\ref{dae} is randomly patched to the tokens or positional embeddings~\cite{he2022masked, devlin2018bert}, and its additive proportion can be adjusted to simulate different levels of noise pollution. The bottleneck design of DAE guarantees the elimination of noise contained in data, as well as enhances the robustness of the feature extractor. The prior knowledge from DAE is that through denoising, the model can be more robust and have better generalizability. 

\noindent\textbf{Momentum Contrast (MoCo):} MoCo is a well-performed contrastive method proposed by He et al.~\cite{he2020momentum} for self-supervised learning. In short, the desired encoder in MoCo is parallelled with a key encoder which encodes the positive and negative pairs on-the-fly. Unlike traditional memory bank design, MoCo adopts the momentum-updated mechanism to renew the target encoder, which can be formulated as:

\begin{equation}
\label{momentum}
\theta_{k} \leftarrow m\theta_{k} + (1-m)\theta{q}, 
\end{equation}
where $m\in [0,1)$ is a momentum coefficient that represents the evolving speed of the paired keys. $\theta_{k},\theta_{q}$ are the parameters of the target encoder and momentum encoder, respectively. Given the optimization method in Function~\ref{momentum}, the goal of MoCo is to minimize InfoNCE~\cite{oord2018representation} as: 

\begin{equation}
\label{infonce}
\mathcal{L}_{q,k^{+},k^{-}}=-\log{\frac{\exp(q\cdot k^{+}/\tau)}{\exp(q\cdot k^{+}/\tau)+\sum_{k^{-}}\exp(q\cdot k^{-}/\tau)}}.
\end{equation}

The optimization result for Function~\ref{infonce} is that the similarity between sample $q$ and positive key $k^{+}$ ($\ell_2(q,k^{+})$) increases, and the dot product of $q$ and negative key $k^{-}$ ($\ell_2(q,k^{-})$) decreases. The prior knowledge from MoCo is that clustering similar sample pairs and alienating dissimilar sample pairs are conducive to supervised learning. 

\noindent\textbf{Simple Contrastive Learning Representation (SimCLR):} Chen et al. introduce a novel contrastive learning algorithm in~\cite{chen2020simple}. Different from previous contrastive learning methods, SinCLR abandons the memory bank and adopts a straightforward method: it randomly samples $N$ samples and leverages stochastic augmentations (i.e., random cropping, random color distortions, and random Gaussian blur) to enlarge the size of this minibatch to $2N$, then it compares the difference between positive and negative samples with: 

\begin{equation}
\label{losscinclr}
\mathcal{L}_{i,j}=-\log{\frac{\exp(sim(z_{i},z_{j})/\tau)}{\sum_{k=1}^{2N}1_{[k\neq i]}\exp((sim(z_{i},z_{j})/\tau)}},
\end{equation}
where $\tau$ denotes a temperature parameter and $sim(z_{i},z_{j})$ is the cosine similarity between $z_{i}$ and $z_{j}$, i.e., $sim(u,v)=u^{T}v/\Vert u\Vert \Vert v\Vert $. 
In SimCLR, the sample combined with its augmented sample is positive pair, while the other $2(N-1)$ augmented samples are treated as negative samples. 
The prior knowledge behind SimCLR is that clustering the data after the composition of multiple data augmentation operations is in favor of yielding effective representations.

\subsection{Attack Scheme}
We conduct our model extraction attack with four steps: 
\label{subsection: attack scheme}

\noindent\textbf{Collecting Proxy Dataset.} To mount our attack, we first gather our proxy dataset $\mathcal{D}_{proxy}$ from the Internet in accordance with the functionality of the MLaaS products. 
Considering the circumstances where users can upload data to the cloud for training customized models, we consider three cases: 1). The attacker has the same training dataset as the victim; 2). The attacker has no knowledge of the training data, but he knows the functionality of API and collects task-related data from the Internet; 3). The attacker is completely blind to the training set and task, he just gathers an OOD surrogate dataset which is task-irrelevant. 

\noindent\textbf{Embedding the Prior Knowledge.} After we gathered the proxy dataset, we first leverage self-supervised learning to train an encoder based on different principles (RS, BAE, DAE, MoCo, SinCLR). 
To improve the robustness of the encoder $f_{enc}$, we adopt multiple data augmentations to transform the input image into five augmented views including RandomCrop, ColorJitter, RandomGrayscale, GaussianBlur, and RandomFlip. 
After embedding the prior knowledge into $f_{enc}$, the attacker replaces the feature extractor of the substitute model $\hat{\mathcal{O}}$ with $f_{enc}$. 
For all SSL methods, we utilize ResNet50 as the backbone to eliminate the effect of different architectures. 
We leave other structures such as Masked Autoencoder~\cite{he2022masked} adopting Vision Transformer~\cite{dosovitskiy2020image} for further exploration. 

\noindent\textbf{Initialing the Start-up Query Dataset.} Different from previous methods which randomly sample points as their start-up query dataset, ours yields the most informative examples based on its prior knowledge learned from the whole $\mathcal{D}_{proxy}$. 
That is, as the $f_{enc}$ estranges the dissimilar samples, the feature vectors of two distinct samples should have diverse patterns that result in low information density. 
This automatic clustering mechanism finds out the uncertain samples according to their natural relation. 

\noindent\textbf{Extracting the Victim Model with Interactions.} After startup, our attack consists of iteratively 1). querying the cloud-based model with sampled informative data, 2). training local substitute model with annotations from the victim, and 3). determining the most informative points for the next round in accordance with their entropy. 
Recall that the prior knowledge of the proxy dataset is pre-compiled into the substitute model, hence the newly selected examples are determined by the prior and posterior knowledge together. 
This paradigm efficiently addresses the overfitting problem when the annotated data is little and has great generalizability when the query budget increases.

\section{Evaluation}
\label{sec:experiments}

\subsection{Experimental Setup}
\label{subsection: experimental setup}

\noindent\textbf{Attack Scenarios:} We conduct our experiments in both experimental and real-world scenarios. 

\textbf{Local Experiments}: To better study the effect of the distribution of proxy dataset and architecture choice of the victim model, we first conduct extensive experiments locally. Strictly following the paradigm of cloud APIs, we first train DNN models with private datasets and then encapsulate them as black boxes. Same to real-world scenarios, the model architecture and hyper-parameters are unknowable to the attacker. We use two widely used benchmark datasets CIFAR10 and STL10 in this scenario: 
\begin{itemize}
    \item \textbf{CIFAR10~\cite{krizhevsky2009learning}}: CIFAR10 is a widely used balanced dataset for object classification tasks. In particular, this dataset consists of 60,000 color $32 \times 32$ images (50,000 training images and 10,000 testing images) from 10 categories including airplane, automobile, bird, cat, deer, dog, frog, horse, ship, and truck. In this case study, we set both the surrogate dataset of the attacker and the training dataset of the victim model to the training set of CIFAR10. 

    \item \textbf{STL10~\cite{coates2011analysis}}: STL10 is a massive dataset for developing self-taught learning algorithms. This dataset contains 5,000 training images and 8,000 testing images from 10 categories including airplane, bird, car, cat, deer, dog, horse, monkey, ship, and truck. In addition, STL10 incorporates an unlabelled subset with a total capacity of 100,000 from classes other than the aforementioned ten classes, such as chicken, snake, beer, and so on. The unlabelled set is out-of-distribution from the training dataset. For this case study, we set the training dataset as a proprietary dataset for the victim and treat the unlabelled dataset as the proxy dataset to mount our attack. 
\end{itemize}

\textbf{Real-world APIs}: We also conduct four case studies (Traffic Recognition, Flower Recognition, NSFW Recognition, and Moderation Recognition) in real-world scenarios. As elucidated in Table~\ref{tab: prediction API platforms and pricing}, the commercialized APIs could be categorized into customization and non-customization. For each condition, we implement our evaluations on two products hosted by popular MLaaS platforms Microsoft and Clarifai, respectively. For customization conditions, a third party can upload his well-labeled dataset to the cloud and the platform will deploy a model which is trained on his data. Then the model owner can open his models' APIs to downstream customers. In this case, the third party has access to the full training dataset while the public or users do not. For another condition, different from the customized MLaaS provider like Microsoft and Google, platforms such as Clarifai and $Face^{++}$ only allow users to query their deployed models and do not support customization. Generally speaking, these APIs usually have a better performance for common tasks while performing poorly on niche tasks. To study the vulnerability of non-customized APIs, we conduct attacks on the Moderation Recognition API and NSFW Moderation API from Clarifai. For both APIs, the training data and model are completely unknown to the public including the attacker. The only operation that an attacker can do is querying the provided APIs and then observe the prediction from these APIs. 

\renewcommand{\arraystretch}{1.2}
\begin{table}[t]
\setlength{\belowcaptionskip}{5pt}
\setlength{\abovecaptionskip}{0.cm}
\scalebox{1.0}{
\begin{tabular}{ccccccc}
\hline
\multirow{3}{*}{Dataset} & \multirow{3}{*}{\begin{tabular}[c]{@{}c@{}}Query\\ Budget\end{tabular}} & \multicolumn{5}{c}{Prior Knowledge}                                                    \\ \cline{3-7} 
                         &                                                                         & \multirow{2}{*}{RS} & \multicolumn{2}{c}{Generative} & \multicolumn{2}{c}{Contrastive} \\ \cmidrule(r){4-5} \cmidrule(r){6-7}
                         &                                                                         &                     & BAE            & DAE           & MoCo               & SimCLR     \\ \hline
\multirow{3}{*}{CIFAR10} & 1k                                                                      & 35.38               & 38.19          & 38.13         & \textbf{71.75}     & 63.60      \\
                         & 2k                                                                      & 40.96               & 41.27          & 45.02         & \textbf{76.40}     & 70.19      \\
                         & 4k                                                                      & 46.54               & 51.72          & 49.57         & \textbf{80.25}     & 79.12      \\ \hline
\multirow{3}{*}{STL10}   & 1k                                                                      & 30.26               & 33.45          & 33.64         & \textbf{62.69}     & 51.64      \\
                         & 2k                                                                      & 34.78               & 36.80          & 38.22         & \textbf{74.48}     & 62.39      \\
                         & 4k                                                                      & 41.25               & 42.63          & 46.00         & \textbf{81.94}     & 71.85      \\ \hline
\end{tabular}
}
\caption{The experimental results of our attack in different scenarios. The CIFAR10 and STL10 experiments are conducted locally to study the impact of IID and OOD. }
\label{table1: ablation exp table}
\vspace{-12pt}
\end{table}
\renewcommand{\arraystretch}{1}

\renewcommand{\arraystretch}{1.1}
\begin{table*}[t]
\setlength{\belowcaptionskip}{5pt}
\setlength{\abovecaptionskip}{0.cm}
\scalebox{1.1}{
\begin{tabular}{ccccccccc}
\hline
\multirow{3}{*}{Platform}  & \multirow{3}{*}{API Type}   & \multirow{3}{*}{Query Budget} & \multicolumn{5}{c}{Prior Knowledge}                                                                    & \multirow{3}{*}{Price (\$)} \\ \cline{4-8}
                           &                             &                               & \multirow{2}{*}{RS} & \multicolumn{2}{c}{Generative Methods} & \multicolumn{2}{c}{Contrastive Methods} &                             \\ \cmidrule(r){5-6} \cmidrule(r){7-8}
                           &                             &                               &                     & BAE           & DAE                    & MoCo               & SimCLR             &                             \\ \hline
\multirow{6}{*}{Microsoft} & \multirow{3}{*}{Traffic}    & 1.00k                         & 27.85               & 16.52         & 29.51                  & \textbf{34.29}     & 24.12              & 2.00                        \\
                           &                             & 2.00k                         & 34.29               & 27.57         & \textbf{42.27}         & 35.11              & 38.54              & 4.00                        \\
                           &                             & 4.00k                         & 41.84               & 42.11         & 45.47                  & 47.21              & \textbf{46.55}     & 8.00                        \\ \cline{2-9} 
                           & \multirow{3}{*}{Flower}     & 0.50k                         & 17.25               & 18.82         & 21.18                  & 17.65              & \textbf{33.53}     & 1.00                        \\
                           &                             & 1.00k                         & 20.00               & 22.35         & 27.65                  & 27.87              & \textbf{38.82}     & 2.00                        \\
                           &                             & 1.50k                         & 22.35               & 28.24         & 32.94                  & 25.88              & \textbf{50.00}     & 3.00                        \\ \hline
\multirow{6}{*}{Clarifai}  & \multirow{3}{*}{Moderation} & 0.24k                         & 73.52               & 76.46         & 78.46                  & 73.92              & \textbf{83.42}     & 0.29                        \\
                           &                             & 0.48k                         & 76.22               & 77.31         & 77.44                  & 74.04              & \textbf{83.78}     & 0.58                        \\
                           &                             & 0.96k                         & 79.74               & 80.34         & 80.16                  & 77.38              & \textbf{85.10}     & 1.15                        \\ \cline{2-9} 
                           & \multirow{3}{*}{NSFW}       & 0.36k                         & 48.50               & 52.10         & 52.00                  & 51.65              & \textbf{84.05}     & 0.43                        \\
                           &                             & 0.72k                         & 50.85               & 53.60         & 74.95                  & 51.95              & \textbf{89.85}     & 0.86                        \\
                           &                             & 1.08k                         & 50.25               & 65.70         & 84.45                  & 58.75              & \textbf{91.50}     & 1.30                        \\ \hline
\end{tabular}
}
\caption{The experimental results of our attack conducted in real-world MLaaS products. The query pricing for Traffic Recognition and Flower Recognition from Microsoft is 2\$ per k queries. For Moderation Recognition API and NSFW Recognition from Clarifai, the query price is 1.2\$/k units. }
\label{table2: ablation exp table}
\end{table*}
\renewcommand{\arraystretch}{1}

\begin{itemize}
    \item \textbf{Traffic Recognition}: The Traffic Sign Recognition API is provided by the Microsoft Azure AI platform, and is used to recognize traffic signs in autonomous driving. We upload the well-labeled GTSRB dataset~\cite{GTSRB} to train the victim model online, and then query this API with samples. GTSRB has a total volume of 39,209 labeled data from 43 types of traffic signs. We split it into a training dataset (31,368 images, 80\% total) and a testing dataset (7,841 images, 20\% total). We set the proprietary training dataset and proxy dataset the same to simulate a condition where the data uploader is malicious and aims to steal the cloud-based model or the private dataset is leaked. 
    
    \item \textbf{Flower Recognition}: Flower Recognition is an API hosted by Azure to recognize the category of the flower. We upload the Flower17 dataset (1,360 points) as the victim model training set. After deploying the victim model, Flower Recognition API outputs a probability distribution among 17 different concepts of flowers. Stealing the model behind the Flower Recognition API is non-trivial as flowers usually have large scale, pose, and light variations. To conduct our attack, we leverage the task-related Flower102 dataset (6,146 points in total)~\cite{flower102} from 102 flower types as our proxy dataset. 
    
    \item \textbf{Moderation Recognition}: The Moderation Recognition API is hosted by Clarifai to analyze the images or videos. It yields the probability scores on the likelihood that the media contains nudity, sexually explicit, otherwise harmful, or abusive user-generated content imagery. The model behind Moderation Recognition API is pre-trained by Clarifai with a proprietary dataset from 5 categories (gore, drug, explicit, suggestive, and safe). To mount our attack, we collect 15,000 images from the Internet for these 5 concepts (3,000 points for each category) within the scope of this service. We also gather another 5,000 images in five classes to evaluate the performance of the substitute models. 

    \item \textbf{NSFW Moderation}: Clarifai Not Safe For Work (NSFW) Moderation API aims to automatically identify whether the content in the images or video is safe for viewing (SFW) or not safe for viewing (NSFW) in American workplaces. This API provides predictions at a coarse granularity, i.e., NSFW or SFW. However, the exact training dataset for this API, the model architecture, and the hyper-parameters adopted to train this model are unknown to the public. To extract this victim model, we collect 5,000 images (2,500 points per category) from the public as the attacker's surrogate dataset, and we also gather another dataset as the testing dataset with a total volume of 2,000. 
\end{itemize}

\noindent\textbf{Evaluation metrics:} Focusing on the fidelity goal, we measure the performance of our attack with metric defined in Formula~\ref{fidelity metric}. To measure the performance of our attack for downstream attacks, we use Attack Success Rate (ASR) to quantify the transferability of adversarial examples generated based on the substitute model.

\subsection{Attack Effectiveness}

\begin{figure*}[t]
\centering
\setlength{\abovecaptionskip}{0.cm}
\includegraphics[width=\textwidth]{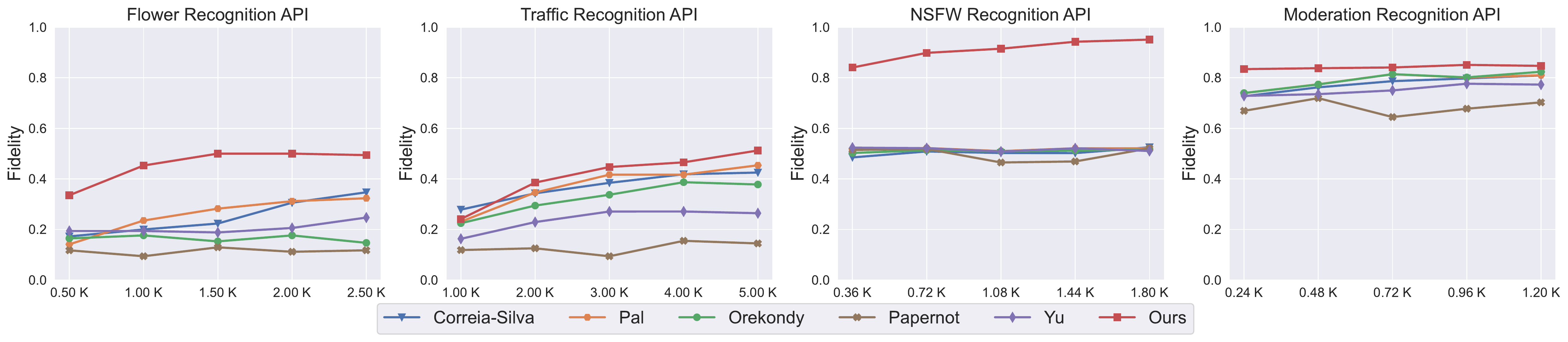}
\caption{Comparison of the fidelity of the stolen model using our attack and other state-of-the-art attacks on real-world MLaaS products.}
\label{Comparison with existing MEAs}
\vspace{-12pt}
\end{figure*}

\begin{figure}[t]
\centering
\setlength{\abovecaptionskip}{0.cm}
\includegraphics[width=\linewidth,scale=1.00]{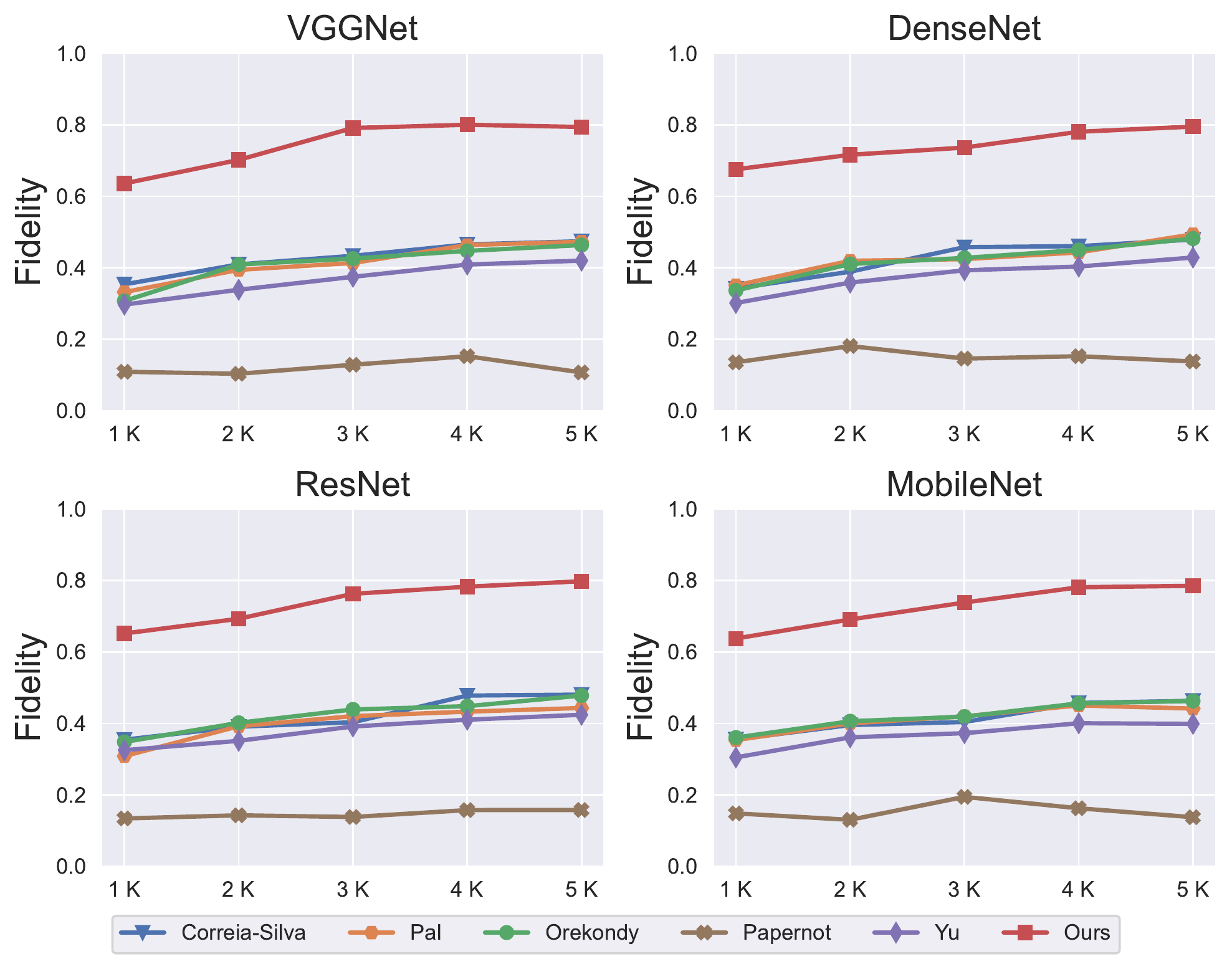}
\caption{Comparison of our attacks with others on IID case. The dataset utilized to conduct attacks is CIFAR10. }
\label{fig:comCIFAR10}
\vspace{-12pt}
\end{figure}

\textbf{IID and OOD Cases:} We first evaluate the effectiveness of our attack on IID (CIFAR10 task) and OOD (STL10 task) cases. We adopt VGGNet13 architecture as the victim model and ResNet50 as the backbone of the substitute model. After training the victim models, we take them as black boxes and open their APIs only. The experimental results are reported in Table~\ref{table1: ablation exp table}. We use five strategies to sample the most informative examples from the proxy dataset. In both IID and OOD cases, the substitute model trained with MoCo achieves the highest fidelity among all query budgets. For the CIFAR10 task, with 1k queries, our substitute model obtains 35.38\% fidelity with RS strategy, 38.19\% fidelity with BAE, 38.13\% fidelity with DAE, 71.75\% fidelity with MoCo, and 63.60\% fidelity with SimCLR. It is worth noting that the MoCo strategy achieves 71.75\% fidelity with 1k queries only, indicating that the effectiveness of our attack is rather high. As the query overhead increases, the agreement between the victim and the substitute also improves. For example, the fidelity obtained by RS raises from 40.96\% to 46.54\% when the query budget increases from 2K to 4K. For the STL10 task, there are similar experimental phenomena to the CIFAR10 task. After querying 4k OOD data, the substitute model achieves 81.94\% fidelity with MoCo and 71.85\% with SimCLR, indicating that our attack is still effective when the distribution of the proxy dataset is different from that of the victim training dataset. 

\noindent\textbf{Case Study 2: Traffic Recognition API. } We upload the GTSRB dataset to the AI platform Azure of Microsoft, and Azure automatically trains and deploys a model on the cloud for us. We conduct model extraction attacks on this API and report the experimental results in Table~\ref{table2: ablation exp table}. As illustrated, with 4.00k queries, our substitute model achieves 41.84\% fidelity with RS strategy, 42.11\% fidelity with BAE strategy, 45.47\% fidelity with DAE strategy, 47.21\% fidelity with MoCo strategy and 46.55\% fidelity with SimCLR strategy.

\noindent\textbf{Case Study 3: Flower Recognition API. } In this case, the cloud-based model is a black box for the attackers, and attackers utilize Flower102 to conduct model extraction attacks. The experimental results of MEAs are shown in Table~\ref{table2: ablation exp table}. As we can see, with 1.50k queries, our attack achieves 22.35\%, 28.24\%, 32.94\%, 25.88\%, and 50.00\% fidelity with RS, BAE, DAE, MoCo, and SimCLR strategies, respectively. Among all sampling strategies, SimCLR achieves the highest fidelity at all query budgets.

\noindent\textbf{Case Study 4: Moderation Recognition API. }We also implement model extraction attacks on the non-customizable models from MLaaS platforms. Here we consider the Moderation Recognition API from Clarifai, which is used to audit the content of images and videos. As demonstrated in Table~\ref{table2: ablation exp table}, with 0.24k queries, our attack achieves 73.52\% fidelity with RS strategy, 76.46\% fidelity with BAE strategy, 78.46\% fidelity with DAE strategy, 73.92\% fidelity with MoCo strategy, and 83.42\% with SimCLR strategy, which indicate that our attack can efficiently steal the Moderation Recognition model with a high agreement. The experimental results show that the SimCLR strategy is better than the other four strategies.

\noindent\textbf{Case Study 5: NSFW Recognition API. }Besides the Moderation Recognition API, we also conduct MEAs on another official API NSFW Recognition from Clarifai. As we can see from Table~~\ref{table2: ablation exp table}, with 1.08k queries and 1.30\$, our substitute model could achieve 91.50\% agreement with the victim model, outperforming the other four strategies, i.e., 50.25\% fidelity with RS, 65.70\% with BAE, 84.45\% with DAE, and 58.75\% with MoCo. This extraction performance is alarming as the attacker extracts a cloud-based model with over 90\% fidelity under a few budgets. 

\begin{figure}[t]
\centering
\setlength{\abovecaptionskip}{0.cm}
\includegraphics[width=\linewidth,scale=1.00]{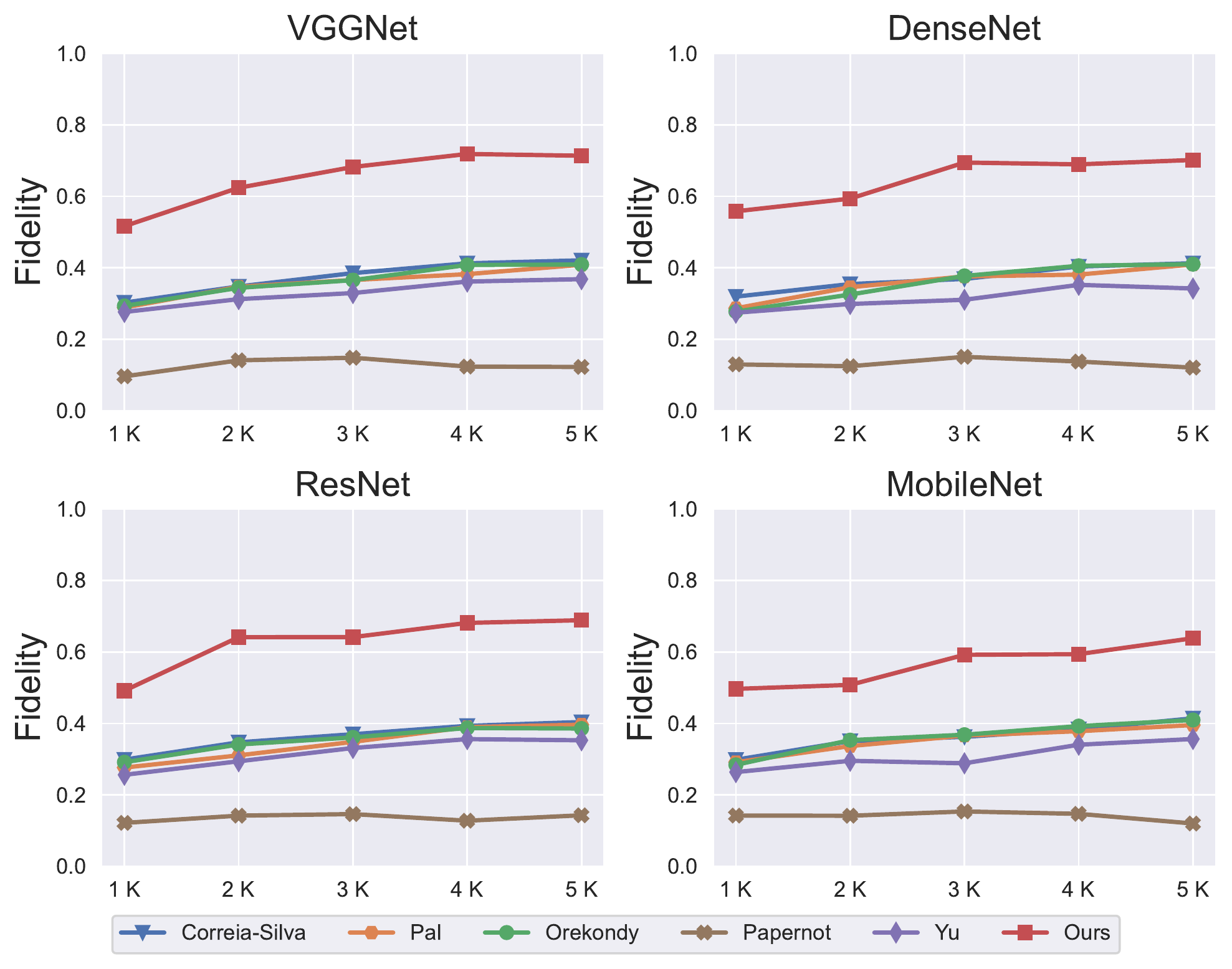}
\caption{Comparison of our attacks with others on OOD case. The dataset utilized to conduct attacks is the unlabeled dataset of STL10.}
\label{fig:comSTL10}
\vspace{-12pt}
\end{figure}

\begin{figure*}[htp]
\centering
\setlength{\abovecaptionskip}{0.cm}
\includegraphics[width=\textwidth]{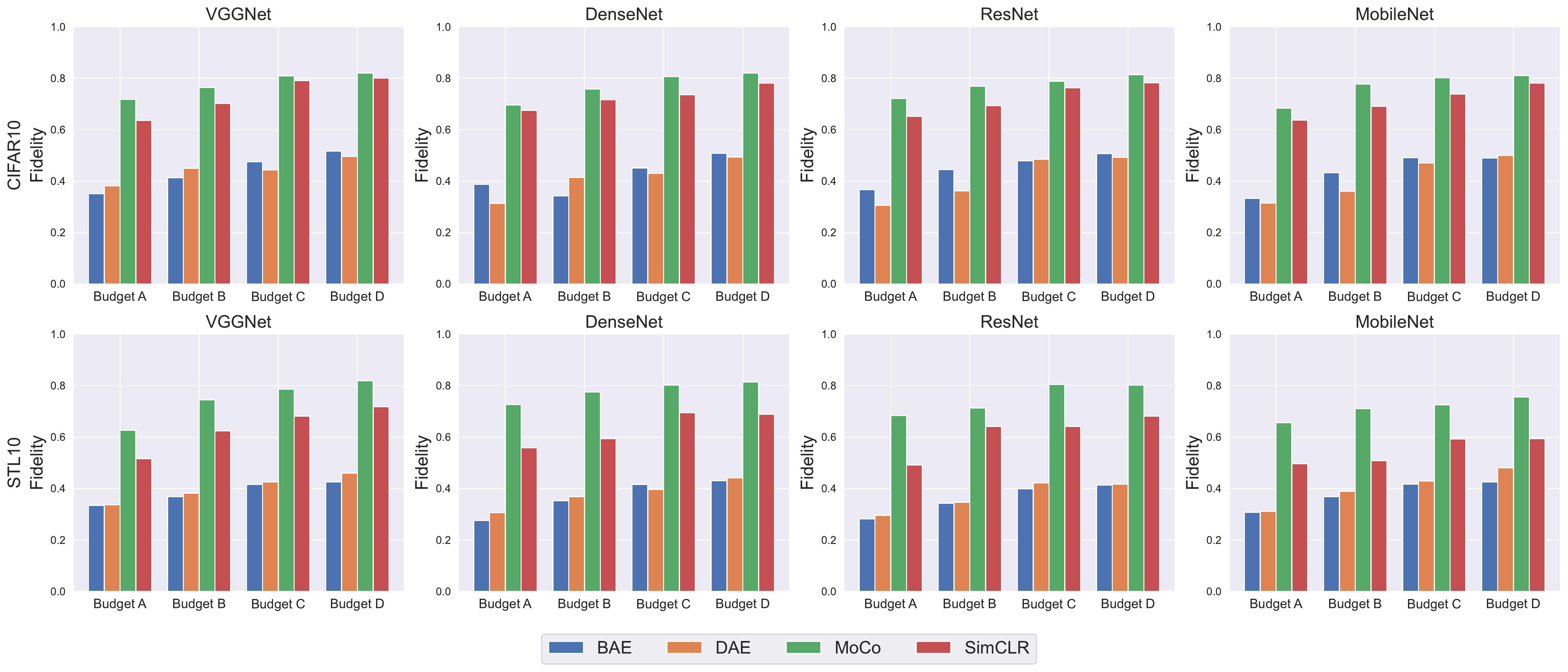}
\caption{Ablation study on the architecture of victim model (Budget A = 1K, Budget B = 2K, Budget C = 3K, Budget D = 4K). }
\label{impact of victim architecture}

\end{figure*}

\subsection{Comparisons with Others}

\textbf{Previous Works.} We compare our attack with five state-of-the-art MEAs, i.e., Correia-Silva attack~\cite{correia2018copycat}, Pal attack~\cite{pal2020activethief}, Orekondy attack~\cite{orekondy2019knockoff}, Papernot attack~\cite{papernot2017practical} and Yu attack~\cite{yu2020cloudleak}. 
Specifically, Correia-Silva et al.~\cite{correia2018copycat} adopts a random strategy to sample query data from the adversary's proxy data $\mathcal{D}_{proxy}$, and then trains a substitute model on these data and the query results from the victim model. Pal et al.~\cite{pal2020activethief} utilize multiple strategies combining active learning to steal the victim model. In this paper, we compare our method with the Uncertainty method proposed in~\cite{pal2020activethief}. Similar to the Pal attack, Orekondy et al.~\cite{orekondy2019knockoff} also use an active sampling strategy to query from the victim model. However, the Orekondy attack leverages reinforcement learning to select query samples. Papernot attack et al.~\cite{papernot2017practical} utilizes the Jacobian Matrix of the substitute model to deploy data augmentation and uses the newly generated data to query the victim model. The FeatureFool (FF) strategy in Yu attack~\cite{yu2020cloudleak} adopts L-BFGS with triplet loss to generate adversarial examples to deduce the decision boundary of the victim model. We reproduce the FF method and compare our attack with it. In our attack, we adopt SimCLR to compile the prior knowledge into the substitute model.

\noindent\textbf{Comparisons in Real-world Scenarios.} Firstly, we compare our attack with the above schemes (Correia-Silva, Pal, Orekondy, Papernot, and Yu) on commercialized APIs including Flower Recognition, Traffic Recognition, NSFW Recognition, and Moderation Recognition API. To ensure fairness, we set the substitute model architectures and hyper-parameters for all attacks to the same. We show the comparison result in Figure~\ref{Comparison with existing MEAs}. As demonstrated, our attack outperforms other attacks for all real-world scenarios. Especially for Flower Recognition and NSFW Recognition case studies, our method surpasses other schemes by a large margin. For example, with 1.80K queries and 2.16 \$, our attack achieves 95.1\% fidelity to the NSFW Recognition API, ahead of the SOTA performance obtained by the Correia-Silva attack.

\noindent\textbf{More Comparisons.} In addition to the real-world scenarios, we also conduct extensive experiments on the distribution of proxy datasets. We compare the efficiency of our attack with other attacks with different query budgets and victim architecture. Different from the CIFAR10 scenario, where the proxy dataset is the same as the victim's training set, the proxy dataset of the STL scenario is OOD. 
As we can see from Figure~\ref{fig:comCIFAR10} and Figure~\ref{fig:comSTL10}, our attack surpasses the existing state-of-the-art attacks by large margins, demonstrating that our scheme is much more efficient. In Figure~\ref{fig:comCIFAR10}, our attack achieves the highest fidelity 79.41\%/79.51\%/79.79\%/78.52\% when the architecture of the victim model is VGGNet13, DenseNet121, ResNet50, and MobileNet V2, respectively. This result illustrates that our attack generalizes across different victim model architectures better than other attacks. Moreover, as the query budget increases from 1000 to 5000, the margins between our attack and others grow, which indicates the efficiency of our attack in all budgets. The same experimental phenomenon also appears in the STL10 scenario, which demonstrates that our attack still outperforms other attacks even when the proxy dataset is completely irrelevant to the victim training set. 

\begin{figure*}[htp]
\centering
\setlength{\abovecaptionskip}{0.cm}
\includegraphics[width=\textwidth]{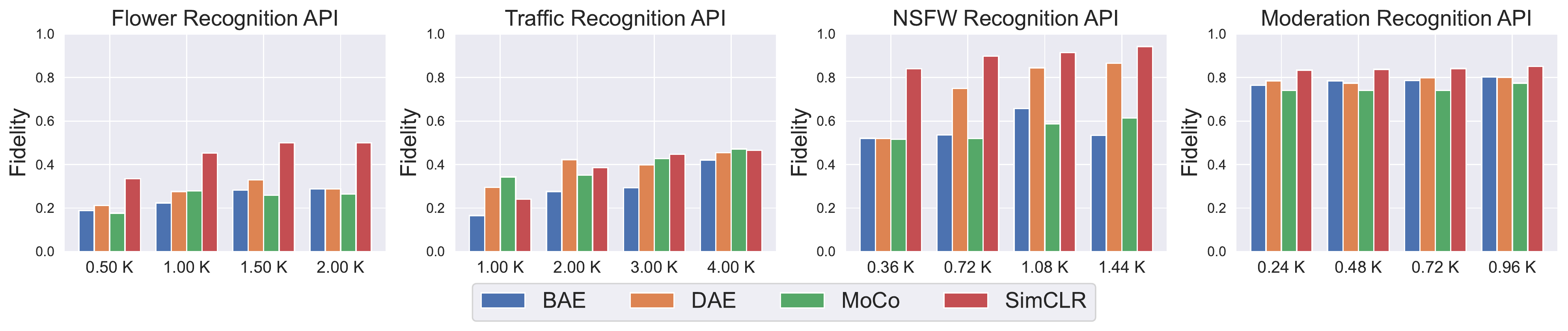}
\caption{Ablation study on the choice of prior knowledge. }
\label{impact of prior knowledge}
\vspace{-12pt}
\end{figure*}

\begin{figure}[t]
\centering
\setlength{\abovecaptionskip}{0.cm}
\includegraphics[width=0.7\linewidth,scale=1.00]{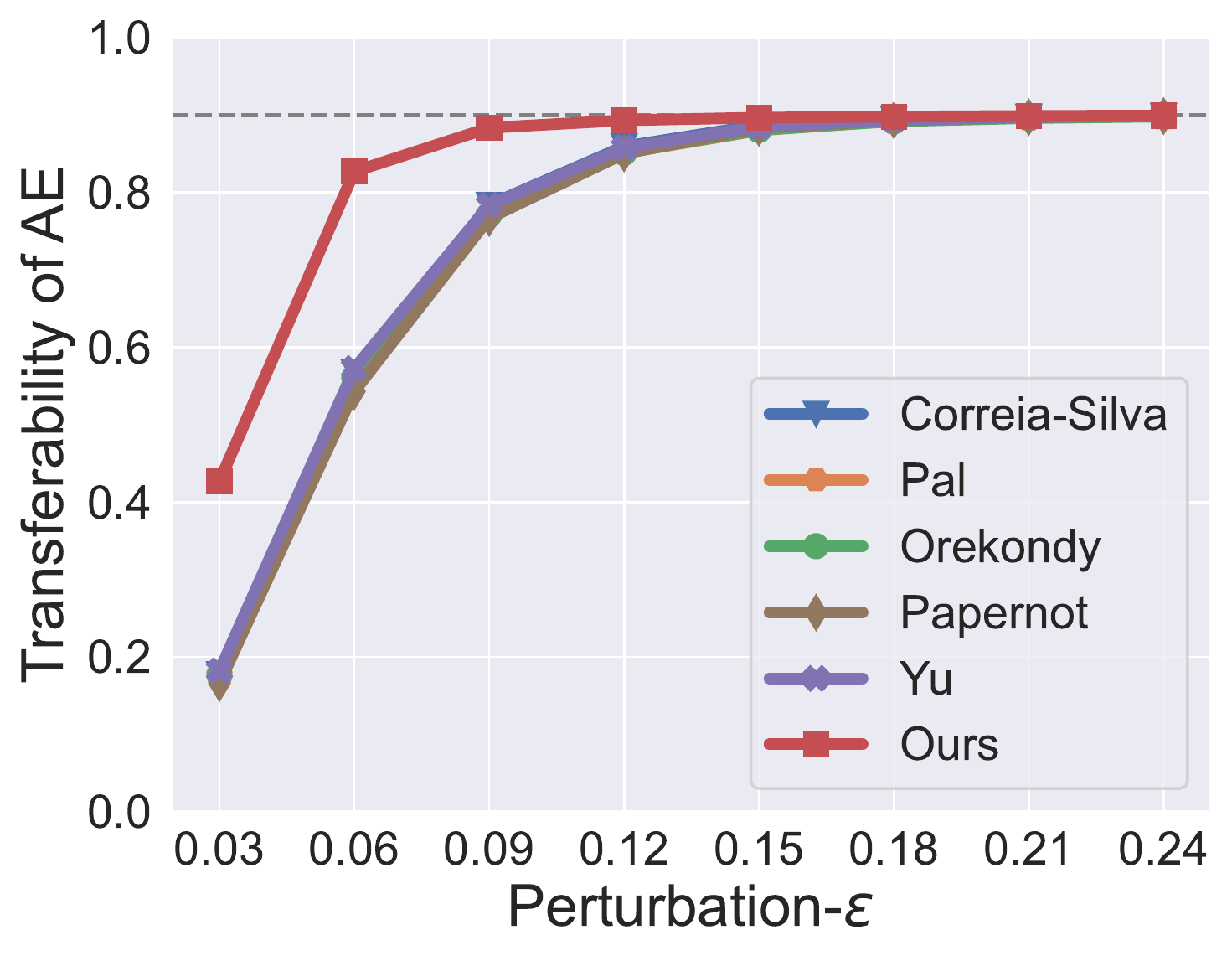}
\caption{Comparison of the transferability of adversarial example generated with different schemes for IID case. In this case, we use the CIFAR10 dataset. }
\label{fig:ae_transferability1}
\vspace{-12pt}
\end{figure}

\subsection{Other Impacts}
In this section, we further study the impacts of the architecture choice of the victim model and prior knowledge. 

\noindent\textbf{Impact of Victim Architecture.} The models with the same architecture are more likely to have a similar performance. Therefore, we explore the impact of the architecture choice of the victim model. Figure~\ref{impact of victim architecture} shows our experimental results when we consider different victim model structures, i.e., VGGNet13, DenseNet121, ResNet50, and MobileNetV3. For the substitute model, we adopt ResNet50 as its architecture. As illustrated, in both CIFAR10 and STL10 tasks, the fidelity performance of our attack is similar. Therefore, our attack generalizes well when the victim model adopts different structures. This is of great significance when considering real-world scenarios where the architecture of the cloud-based model is unknown to the attacker. 

\noindent\textbf{Impact of Prior Knowledge.} The principle of self-supervised learning determines the prior knowledge embedded in the substitute model. Roughly, self-supervised learning can be categorized into the generative method and the contrastive method. We study the impact of different self-supervised learning methods and report their results in Figure~\ref{impact of prior knowledge}. Getting rid of multifarious settings, we conduct this ablation study in four real-world APIs as introduced before. We find that for Flower Recognition, NSFW Recognition, and Moderation Recognition API, SimCLR achieves the highest extraction fidelity at all costs, while another contrastive method MoCo obtains a similar performance with the generative methods BAE and DAE or even lags behind. We suspect the reason is that the self-taught process of SimCLR is more like supervised learning. Therefore, the prior knowledge it learns is transferable to the model extraction attack.

\subsection{Transferability of Adversarial Example}
This section evaluates the transferability of the adversarial example (AE) generated based on the stolen substitute model. Up to date, the AE generation of existing adversarial example attacks mainly depends on the gradient of the victim model. However, in practice, it is an extremely high requirement as the victim models are black boxes for the attackers most of the time. Considering such a predicament, Papernot et al.~\cite{papernot2017practical} propose a new attack paradigm where the attacker first utilizes Jacobian Matrix to steal a cloud-based model, and then leverages the substitute model to craft AEs to make the victim misclassify~\cite{pal2020activethief}. Intuitively, a substitute model with high fidelity should have a similar decision boundary to the target model. Therefore, the transferability of AE generated on the basis of a substitute model should be rather high. 

In this paper, we utilize the Fast Gradient Sign Method (FGSM)~\cite{szegedy2013intriguing} to generate adversarial examples, and then measure their transferability to the victim model. FGSM~\cite{goodfellow2014explaining} is a one-step AE-generating attack that adopts the gradient ascent method to optimize the adversarial example. Formally, the FGSM generates AE in accordance with:

\begin{equation}
\label{fgsm}
x = x+\epsilon sign (\nabla_{x}\mathcal{L}(\theta,x,y)), 
\end{equation}
where $\epsilon$ determines the magnitude of disturbance. Here, we define the transferability of AE as the percentage of the crafted AEs that are misclassified by the victim. We evaluate the performance of the generated AEs at the rates of $\epsilon=0.03, 0.06, 0.09, 0.12, 0.15, 0.18, 0.24$.

\begin{figure}[t]
\centering
\setlength{\abovecaptionskip}{0.cm}
\includegraphics[width=0.7\linewidth,scale=1.00]{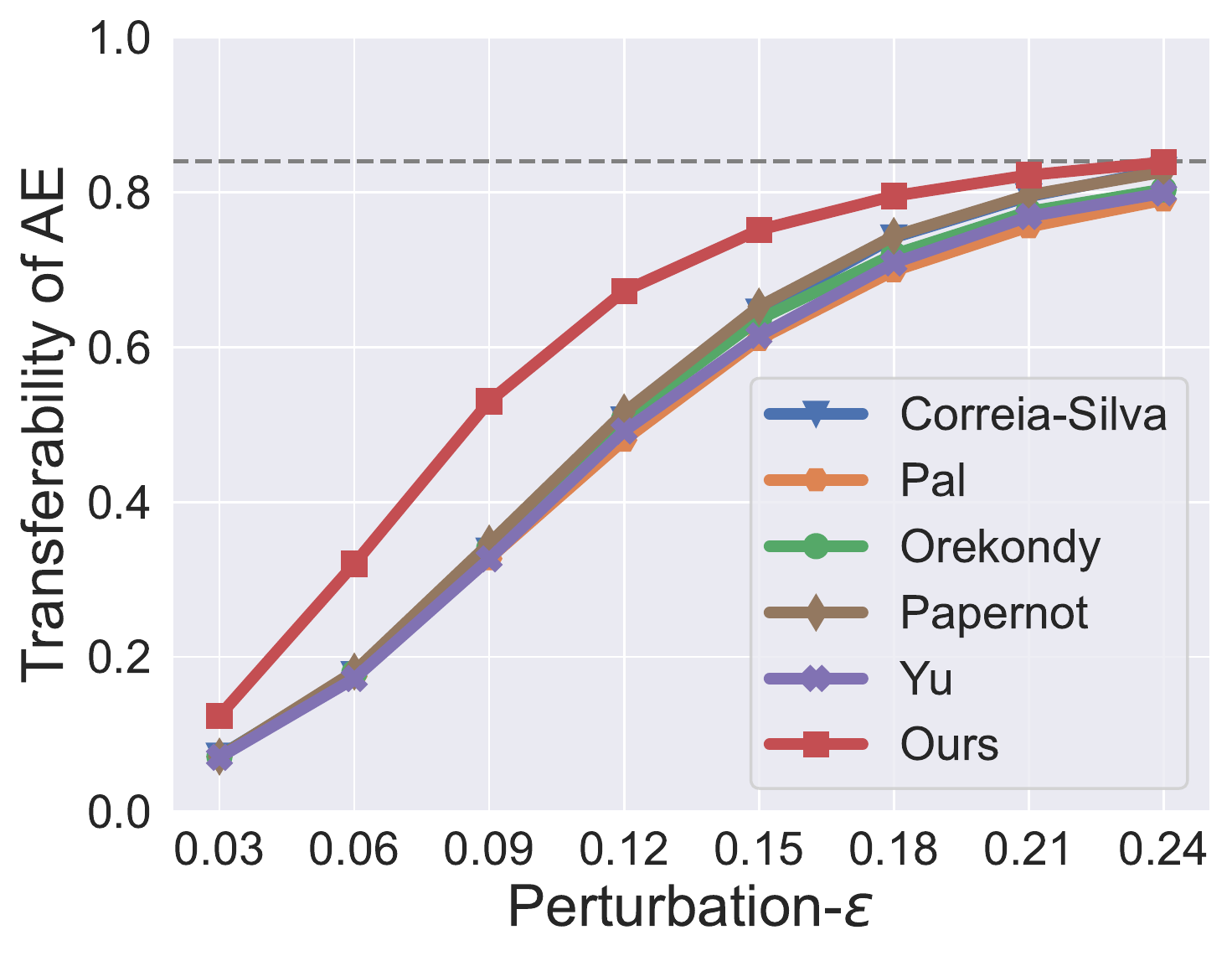}
\caption{Comparison of the transferability of adversarial example generated with different schemes for OOD case. In this case, we use the STL10 dataset. }
\label{fig:ae_transferability2}
\vspace{-12pt}
\end{figure}

\noindent\textbf{Transferability Comparisons.} We conduct model extraction attacks and adversarial example attacks on IID and OOD cases. In the first stage of the attack, i.e., MEA, we set the query budget as 2,000 for both cases. After extracting the victim model, we leverage the substitute model to generate adversarial examples with the FGSM attack for the downstream attack. As demonstrated in Figure~\ref{fig:ae_transferability1}, for the IID case, the AE generated by our method has better transferability across a wide range of perturbations. Our method outperforms other schemes by a large margin, especially when the perturbation is small. For example, our attack achieves 82.70\% ASR when the perturbation size is only 0.06. 
Moreover, we also conduct adversarial example attacks with an OOD proxy dataset. As we can see from Figure~\ref{fig:ae_transferability2}, our attack maintains the highest transferability as the perturbation $\epsilon$ increases from 0.03 to 0.24. Our attack is clearly ahead of other attacks when the perturbation is about 0.09 in terms of transferability. Another significant superiority of our attack is that when achieving the same ASR, our attack requires a much smaller perturbation, indicating that our AEs are harder to be visually detected. Therefore, our model extraction attack is more conducive to downstream attacks than the existing MEAs.

\noindent\textbf{Impact of SSL.} Table~\ref{table: ae_cifarl10} and \ref{table: ae_stl10} show the ASR of AEs generated from the stolen model with different strategies. As we can see from Table~\ref{table: ae_cifarl10}, the substitute model extracted with SimCLR achieves the highest ASR for $\epsilon=0.06,0.09,0.12,0.15$. Another contrastive method MoCo obtains the highest ASR when the perturbation is 0.03. Totally speaking, both the AEs crafted by generative methods and contrastive methods achieve high transferability when the $\epsilon$ is rather high. However, the contrastive methods behave better at a small perturbation. For the OOD case, there is a similar phenomenon to the IID case. It is noteworthy that MoCo dominates at most perturbations except $\epsilon=0.03$ as shown in Table~\ref{table: ae_stl10}.

\renewcommand{\arraystretch}{1.2}
\begin{table}[]
\setlength{\belowcaptionskip}{5pt}
\setlength{\abovecaptionskip}{0.cm}
\begin{tabular}{c|ccccc}
\hline
\multirow{2}{*}{} & \multicolumn{5}{c}{FGSM ASR (\%)}                                                       \\
                  &$\epsilon=0.03$ &$\epsilon=0.06$ &$\epsilon=0.09$ &$\epsilon=0.12$ &$\epsilon=0.15$ \\ \hline
BAE               & 18.44          & 58.61          & 79.40          & 86.37          & 88.71          \\
DAE               & 16.77          & 50.81          & 72.97          & 82.20          & 86.49          \\
MoCo              & \textbf{43.99}          & 82.64          & 88.23          & 89.04          & 89.35          \\
SimCLR            & 42.69 & \textbf{82.70} & \textbf{88.35} & \textbf{89.32} & \textbf{89.66} \\ \hline
\end{tabular}
\caption{The transferability of AE generated by the substitute model extracted with different prior knowledge in IID case. The victim training dataset and attacker proxy dataset are CIFAR10.}
\label{table: ae_cifarl10}
\vspace{-12pt}
\end{table}
\renewcommand{\arraystretch}{1}

\section{Discussion}

\renewcommand{\arraystretch}{1.2}
\begin{table}[]
\setlength{\belowcaptionskip}{5pt}
\setlength{\abovecaptionskip}{0.cm}
\begin{tabular}{c|ccccc}
\hline
\multirow{2}{*}{} & \multicolumn{5}{c}{FGSM ASR (\%)}                                                       \\
                  &$\epsilon=0.03$ &$\epsilon=0.06$ &$\epsilon=0.09$ &$\epsilon=0.12$ &$\epsilon=0.15$    \\ \hline
BAE               & 7.77           & 19.98          & 40.59          & 60.51          & 73.78          \\
DAE               & 7.91           & 19.19          & 36.66          & 53.68          & 66.68          \\
MoCo              & 12.13          & \textbf{32.81} & \textbf{53.40} & \textbf{68.02} & \textbf{77.19} \\
SimCLR            & \textbf{12.39} & 32.02          & 53.01          & 67.26          & 75.16          \\ \hline
\end{tabular}
\caption{The transferability of AE generated by the substitute model extracted with different prior knowledge in the OOD case. The victim training dataset is the training set from STL10, and the proxy dataset is the unlabeled set from STL10.}
\label{table: ae_stl10}
\end{table}
\renewcommand{\arraystretch}{1}

\subsection{Defenses}
\label{defense}
To defend existing MEAs, there is a variety of works providing countermeasures. We give a detailed analysis of these defenses and the possible effect on our attack. 

\noindent\textbf{Detect Malicious Queries.} A simple and straightforward approach to defend MEAs is to analyze the adversary's query examples and block her/his account once being determined to be malicious by the platform. 
For example, in this line, PRADA \cite{juuti2019prada} proposes a  detection method, which studies the distribution of consecutive query samples and sounds the alarm if the distribution is abnormal from benign behavior. 
SEAT \cite{zhang2021seat} presents a similarity encoder to measure the context of query samples' resemblance, and then decides whether to terminate the user's account based on the similarity results. 
Besides, other works such as \cite{pal2021stateful, kesarwani2018model, atli2020extraction} also explore several detection-based strategies.
These defenses could be the most likely countermeasure for attacks with adversarial examples including ours. 

\noindent\textbf{Obstruct Attacker's Acquisition.} Different from the above detection-based solutions, obstruction-based strategies attempt to slow down the adversary's process of obtaining useful information from the victim's API. 
For example, Dziedzic et al. \cite{dziedzic2022increasing} impede MEAs with a calibrated proof-of-work strategy, in which users are required to complete a proof-of-work before they can obtain the model’s predictions. 
This increases the difficulty for adversaries to get the victim's prediction results. 
However, this method is not effective to defend our attack since we only consume a small query budget. 

\noindent\textbf{Perturb Adversary's Training.} Another angle of defense is to limit the information leakage of each query such that adversaries could not obtain useful information to train a substitute model. 
To this end, lots of works propose perturbation-based methods. 
For example, Zheng et al. \cite{zheng2019bdpl} leverage differential privacy to perturb prediction responses near the decision boundary. 
Kariyappa et al. \cite{kariyappa2020defending} add adaptive misinformation into OOD queries to defend adversarial examples-based MEAs. 
However, as this defense needs to modify the prediction of the query sample, it would bear a loss of accuracy. Therefore, it is necessary to consider a better trade-off between prediction accuracy and defensive effect. 

\noindent\textbf{Discriminate Intellectual Property.} Besides the above defenses, some works also propose methods for protecting the model's IP, which can be regarded as another defense for MEAs. 
Generally speaking, these methods protect a model's IP via planting a watermark into the model such that the authentic owner could declare possession of the model~\cite{adi2018turning, zhang2018protecting}. 
For example, Jia et al. \cite{jia2021entangled} propose a watermark-based method, called EWE, to guarantee that the watermark cannot be removed even when retraining. 
However, as the cultivation of a watermark would affect the accuracy of the model, the defender needs to sacrifice the performance to achieve higher security. 
\subsection{Future Work}
\label{future work}
\noindent\textbf{Further exploration of prior knowledge to MEAs.} Our attack introduces self-supervised learning to the model extraction attack. As a hot topic, abundant SSL algorithms are proposed and proved to be efficient when the training data is few such as BYOL~\cite{grill2020bootstrap} and SimSiam~\cite{chen2021exploring}. Adopting the general scheme proposed in this paper, the existing SSL methods can replace the module of prior knowledge learning directly. Moreover, another potential direction is to explore the likelihood of combining unsupervised learning with MEAs. Unsupervised learning clusters unlabeled data in accordance with the distance between samples. In this case, the samples located at the border of clusters are more likely to be sorted as another category. 

\noindent\textbf{Extension to more API types.} During our investigation of commercialized APIs, we find that there are some other kinds of MLaaS products, such as Body Outlining Model\footnote{https://www.faceplusplus.com/body-outlining/} from $Face^{++}$ and Audio Transcription Model\footnote{https://clarifai.com/facebook/asr/models/asr-wav2vec2-base-960h-english} from Clarifai are not studied in previous works. However, there is luxuriant commercial value contained in those APIs thus worth considering their security problem. Due to the differences in modal and task, the existing model extraction attacks may not be used to steal the functionality of those commercialized APIs. An interesting direction is to explore the vulnerability of these models. 

\subsection{Responsible exposure}
\label{responsible exposure}
Before we conduct our attacks on those APIs, we informed the MLaaS providers Microsoft and Clarifai. After we have implemented MEAs, we report the attack effect to them and provide them with some useful countermeasures to protect the IP of cloud-based models.

\section{Conclusion}

In this work, we propose a novel and efficient fidelity-oriented extraction framework to steal the cloud-based model with prior knowledge. 
We pre-compile the prior knowledge of the proxy dataset into the substitute model and explore the utilization efficiency of unlabeled data to the extreme. 
Different from existing MEA attacks, which heavily rely on the posterior knowledge of victim models and hence result in generalization error and overfitting problems under query budget limitations, we combine the prior and posterior knowledge together achieving high efficiency in large-scale cloud-based model stealing attacks. 
The experimental results demonstrate that our attack achieves remarkable extraction performance and surpasses other methods by a gargantuan margin. 
We hope this work could inspire the study of introducing more prior knowledge to the model stealing attacks and raise the alarm about the security of MLaaS.

\bibliographystyle{ACM-Reference-Format}
\bibliography{acmart.bib}

\appendix

\end{document}